\documentstyle[11pt]{article}
\newcommand{\be}{\begin{equation}}
\newcommand{\ee}{\end{equation}}
\newcommand{\bea}{\begin{array}}
\newcommand{\eea}{\end{array}}

\title{SOME SURVEY REMARKS ON\\WHITHAM THEORY AND EM DUALITY}
\author{ROBERT CARROLL\\\footnotesize Mathematics Department, 
University of Illinois,
Urbana, IL 61801\\\footnotesize email:  rcarroll@symcom.math.uiuc.edu}

\date{\begin{minipage}{150mm}\noindent\footnotesize
\end{minipage}}

\begin{document}

\bibliographystyle{plain}
\maketitle



\begin{center}
\section{INTRODUCTION}
\end{center}

\normalsize
\renewcommand{\theequation}{1.\arabic{equation}}\setcounter{equation}{0}

In the last few years there has been enormous progress in string theory,
connected with various ideas of duality.  Similarly in $N=2$
supersymmetric (susy) gauge theory the idea of electromagnetic (EM) duality 
emerged in 1994 with the work of Seiberg-Witten \cite{sb}.  The
mathematical consequences of such work have been no less profound than
the physical import but a survey is not contemplated here.  One theme
which has emerged however involves the interaction of ideas of 
integrability with matrix models, topological field theory (TFT),
EM duality, Landau-Ginsburg (LG) models, etc.  This is perhaps not
surprising in that integrability involving invariant tori etc. seems
fundamental in quantization itself but the interaction here is most
explicit and inevitable.  Without going into a discussion of $N=2$ and
$N=1$ susy let us simply say that the low energy effective action for an
$N=2$ susy Yang-Mills (YM) theory with gauge group $SU(2)$ for example
is described by a holomorphic prepotential ${\cal F}$ (see \cite
{bc,cd,sb}
for a discussion and \cite{fh} for susy - we will not give extensive 
references to the physics here).  This theory has a scalar potential
$V(\phi)=0$ in the vacuum and there are generally nonvanishing $\phi$ for
which this occurs.  Setting $\phi=(1/2)a\sigma_3\,\,(\sigma_3=
\left(
\scriptsize
\begin{array}{cc}
1 & 0\\
0 & -1
\end{array}
\normalsize
\right)$) 
and $u=<Tr\phi^2>=(1/2)a^2$ ($<\,\,\,>\sim$
vacuum expectation value or vev, where $\sim$ is used for ``corresponds to"
or sometimes ``is asymptotic to"), the complex parameter $u$ 
labels inequivalent vacua and the manifold of gauge inequivalent vacua
is called the moduli space ${\cal M}$ of the theory (here $u$ is a
coordinate on ${\cal M}\sim{\bf C}$ with punctures or singularities).
A metric on ${\cal M}$ is given locally by $ds^2=
\Im{\cal F}''(a)dad\bar{a}=\Im\tau(a)dad\bar{a}$ 
where $\tau(a)\sim (\theta/2\pi)+(4\pi i/g^2)$ is the complex coupling
constant.  One writes $a^D=(\partial{\cal F}/\partial a$ (with
$<\phi^D>=(1/2)a^D\sigma_3$) and the superscript $D$ indicates duality
here.  There is a Legendre transformation connecting dual superfields, etc.
and $ds^2$ becomes $ds^2=\Im da^Dd\bar{a}$.  On ${\cal M}$ physical arguments
show that $a(u),\,\,a^D(u)$ are single valued except for monodromies around
say $(u_0,\,-u_0,\,\infty$) (e.g. $(a^D,a)^T\to M_{\infty}(a^D,a)^T$ for
$M_{\infty}=\left(
\scriptsize
\begin{array}{cc}
-1 & 2\\
0 & -1
\end{array}
\normalsize
\right)$) and one picks $u_0=1$ by renormalization arguments.  This leads
to a realization of the monodromies by a differential equation $(\spadesuit)
\,\,(-D_z^2+V(z))\vec{\psi}(z)=0,\,\,\vec{\psi}=((i/2)a,-ia^D)^T$ and this
can be rewritten as a hypergeometric equation from which one obtains
\be
a_D(u)=i\psi_2(u)=i\frac{u-1}{2}F(\frac{1}{2},\frac{1}{2},2,\frac{1-u}{2})=
\frac{\sqrt{2}}{\pi}\int_1^u\frac{dx\sqrt{x-u}}{\sqrt{x^2-1}};
\label{A}
\ee
$$a(u)=-2i\psi_1(u)=\sqrt{2}(u+1)^{\frac{1}{2}}F(-\frac{1}{2},
\frac{1}{2},1,\frac{2}{u+1})
=\frac{\sqrt{2}}{\pi}\int_{-1}^1\frac{dx\sqrt{x-u}}{\sqrt{x^2-1}}$$

\newpage
\topmargin=-26mm
\noindent
One can eliminate $u$ to get e.g. $a^D(a)$.  The integral formulas in
(\ref{A}) are easily recognized as arising from an elliptic curve
(take branch cuts $(-1,1)$ and $(u,\infty)$ in ${\bf C}$ and construct
a two sheeted Riemann surface of genus one (torus)).  Take
homology cycles $A\sim$ cycle $1\to u$ on one sheet and $u\to 1$ on the
other, with $B\sim$ cycle around $(-1,1)$ and define now
\be
a=\oint_Ad\lambda;\,\,a^D=\oint_Bd\lambda;\,\,d\lambda=
\frac{1}{\pi\sqrt{2}}\frac{\sqrt{z-u}}{\sqrt{z^2-1}}dz
\label{B}
\ee
This differential $d\lambda=d\lambda_{SW}$ is the Seiberg-Witten (SW)
differential and the integrals in (\ref{B}) are period integrals which
satisfy $(\spadesuit)$ under the label of Picard-Fuchs (PF) equations.
The correct monodromies are obtained directly (cf. also \cite{tc} for
an interesting use of monodromy ideas).
We will indicate how all
this is related to integrable systems and the Whitham equations.

\begin{center}\footnotesize
\section{WHITHAM EQUATIONS}
\end{center}
\renewcommand{\theequation}{2.\arabic{equation}}\setcounter{equation}{0}

\normalsize

We will work in the context of algebro-geometric solutions of integrable
partial differential equations (PDE) which involves the following 
ingredients (cf. \cite{ba,cb,da}).
Take an arbitrary Riemann surface $\Sigma$
of genus $g$, pick a point $Q$ and a local variable $1/k$ near $Q$
such that $k(Q) = \infty$, and, for illustration, take $q(k)=kx+k^2y+k^3t$. 
Let $D = P_1 + \cdots + P_g$ be a non-special
divisor of degree $g$ and write $\psi$ for the (unique up to a constant
multiplier by virtue of the Riemann-Roch theorem)
Baker-Akhiezer (BA) function characterized by the properties
that $\psi$ is meromorphic on $\Sigma$ except for $Q$ where $\psi
(P)exp(-q(k))$ is analytic 
and (*) $\psi\sim exp(q(k))[1 + \sum_1^{\infty}(\xi_j/
k^j)]$ near $Q$; on $\Sigma/Q,\,\,\psi$ has poles at the $P_i$.
In fact $\psi$ can be taken in the form ($P\in\Sigma,\,\,
P_0\not= Q$)
\be
\psi(x,y,t,P) =  exp[\int^P_{P_0}(xd\Omega^1 + yd\Omega^2 + td\Omega^3)]
\cdot\frac{\Theta({\cal A}(P) + xU + yV + tW + z_0)}{\Theta({\cal A}
(P) + z_0)}
\label{psi}
\ee
where $d\Omega^1 = dk + \cdots,\,\,d\Omega^2 = d(k^2) + \cdots,\,\,
d\Omega^3 = d(k^3) + \cdots, U_j = \int_{B_j}d\Omega^1,\,\,V_j = \int_
{B_j}d\Omega^2,\,\,W_j = \int_{B_j}d\Omega^3\,\,(j = 1,\cdots,g),\,\,z_0
= -{\cal A}(D) - K$, and $\Theta$ is the Riemann theta function.
Here the
$d\Omega_j$ are meromorphic differentials of second kind normalized via
$\int_{A_k}d\Omega_j = 0\,\,(A_j,\,B_j$ are canonical homology cycles)
and we note that $xd\Omega^1 + yd\Omega^2 + td\Omega^3\sim
dq(k)$ normalized;
${\cal A}$ is the Abel-Jacobi map (${\cal A}(P) = \int^Pd\omega_k$ where
the $d\omega_k$ are normalized holomorphic differentials, $k = 1,\cdots,g,
\,\,\int_{A_j}d\omega_k = \delta_{jk}$), and $K = (K_j)\sim$ Riemann
constants ($2K = -{\cal A}(K_{\Sigma})$ where $K_{\sigma}$ is the
canonical class of $\Sigma\sim$ equivalence class of meromorphic 
differentials) so $\Theta({\cal A}(P) + z_0)$ has exactly $g$ zeros
(or vanishes identically).  The paths of integration are to be the
same in computing $\int_{P_0}^Pd\Omega^i$ or ${\cal A}(P)$ and it is
shown in \cite{ba,ca,da} that $\psi$ is well defined (i.e. path independent).  
Then the $\xi_j$ in (*) can be computed
formally and one determines Lax operators $L$ and $A$ such that
$\partial_y\psi = L\psi$ with $\partial_t\psi = A\psi$.  Indeed, given
the $\xi_j$ write $u = -2\partial_x\xi_1$ with $w = 3\xi_1\partial_x\xi_1
-3\partial^2_x\xi_1 - 3\partial_x\xi_2$.  Then formally, near $Q$, one
has $(-\partial_y + \partial_x^2 + u)\psi = O(1/k)exp(q)$ and 
$(-\partial_t + \partial^3_x + (3/2)u\partial_x + w)\psi = O(1/k)exp(q)$
(i.e. this choice of $u,\,w$ makes the coefficients of $k^nexp(q)$ vanish
for $n = 0,1,2,3$).  Now define $L = \partial_x^2 + u$ and $A = \partial^3_x
+ (3/2)u\partial_x + w$ so $\partial_y\psi = L\psi$ and $\partial_t\psi
= A\psi$.  This follows from the uniqueness of BA functions with the same
essential singularity and pole divisors (Riemann-Roch). 
Then we have, via compatibility
$L_t - A_y = [A,L]$, a KP equation $(3/4)u_{yy} = \partial_x[u_t
-(1/4)(6uu_x + u_{xxx})]$ and therefore such KP equations 
arise automatically from a Riemann surface and are parametrized
by nonspecial divisors or equivalently by points in general position
on the Jacobian variety $J(\Sigma)$.
The flow variables $x,y,t$ arise via
$q(k)$ and then miraculously reappear in the theta function via
$xU+yV+tW$; thus the Riemann surface itself contributes to establish
these as linear flow variables on the Jacobian and in a certain sense
defines the flow variables.
The pole positions
$P_i$ do not vary with $x,y,t$ and $(\dagger)\,\,
u = 2\partial^2_x log\Theta(xU + yV  + tW + z_0) + c$ 
exhibits $\Theta$ as a tau function.
\\[3mm]\indent
We recall also that
a divisor $D^{*}$ of degree $g$ is dual to $D$ (relative to $Q$) if
$D + D^{*}$ is the null divisor of a meromorphic differential $d\hat{\Omega}
= dk + (\beta/k^2)dk + \cdots$ with a double pole at $Q$ (look at
$\zeta = 1/k$ to recognize the double pole).  Thus 
$D + D^{*} -2Q\sim K_{\Sigma}$ so ${\cal A}(D^{*}) - {\cal A}(Q) + K =
-[{\cal A}(D) - {\cal A}(Q) + K]$.  One can define then a function
$\psi^{*}(x,y,t,P) = exp(-kx-k^2y-k^3t)[1 + \xi_1^{*}/k) + \cdots]$
based on $D^{*}$ (dual BA function)
and a differential $d\hat{\Omega}$ with zero divisor $D+D^*$, such that
$\phi = \psi\psi^{*}d\hat{\Omega}$ is
meromorphic, having for poles only
a double pole at $Q$ (the zeros of $d\hat{\Omega}$ cancel
the poles of $\psi\psi^{*}$).  Thus $\psi\psi^*d\hat{\Omega}\sim \psi\psi^*(1+
(\beta/k^2+\cdots)dk$ is meromorphic with a second order pole at $\infty$,
and no other poles.  
For $L^{*} = L$ and $A^{*} = -A + 2w
-(3/2)u_x$ one has then $(\partial_y + L^{*})\psi^{*} = 0$ and
$(\partial_t + A^{*})\psi^{*} = 0$.  Now  
the prescription above seems to specify for $\psi^*$
($\vec{U}=xU+yV+tW,\,\,z_0^* =
-{\cal A}(D^*)-K$)
\be
\psi^*\sim e^{-\int^P_{P_o}(xd\Omega^1+yd\Omega^2+td\Omega^3)}
\cdot\frac{\Theta({\cal A}(P)-\vec{U}+z_0^*)}
{\Theta({\cal A}(P)+z_0^*)}
\label{star}
\ee
In any event the message here is that for any Riemann surface 
$\Sigma$ one can
produce a BA function $\psi$ with assigned flow variables $x,y,t,\cdots$
and this $\psi$ gives rise to a 
(nonlinear) KP equation with solution $u$ linearized
on the Jacobian $J(\Sigma)$.
\\[3mm]\indent
One knows that hyperelliptic curves play a special role in the theory
of algebraic curves and Riemann surfaces.  For 
hyperelliptic Riemann surfaces one can pick
any $2g+2$ points $\lambda_j\in {\bf P}^1$ and there will be a unique
hyperelliptic curve $\Sigma_g$ with a 2-fold map $f:\,\Sigma_g\to{\bf P}^1$ 
having branch locus $B=\{\lambda_j\}$.  Since any 3 points $\lambda_i,\,
\lambda_j,\,\lambda_k$ can be sent to $0,\,1,\,\infty$ by an automorphism
of ${\bf P}^1$ the general hyperelliptic surface of genus $g$ can be
described by $(2g+2)-3=2g-1$ points on ${\bf P}^1$.  Since $f$ is unique
up to an automorphism of ${\bf P}^1$ any hyperelliptic $\Sigma_g$ corresponds
to only finitely many such collections of $2g-1$ points so locally there
are $2g-1$ (moduli) parameters.  Since the moduli space of algebraic
curves has dimension $3g-3$ one sees that for $g\geq 3$ the generic 
Riemann surface is nonhyperelliptic whereas for $g=2$ all Riemann
surfaces are hyperelliptic (with 3 moduli).  For $g=1$ we have tori
or elliptic curves with one modulus $\tau$ and $g=0$ corresponds to
${\bf P}^1$.  In many papers on soliton mathematics and integrable systems
one takes real distinct branch points $\lambda_j,\,\,1\leq j\leq 2g+1$, and 
$\infty$, with $\lambda_1<\lambda_2<\cdots<\lambda_{2g+1}<\infty$ and
$\mu^2=\prod_1^{2g+1}(\lambda-\lambda_j)=P_{2g+1}(\lambda,\lambda_j)$
as the defining equation for $\Sigma_g$.  Evidently one could choose
$\lambda_1=0,\,\,\lambda_2=1$ in addition so for $g=1$ we could use
$0<1<u<\infty$ for a familiar parametrization with elliptic integrals,
etc.  One can take $d\lambda/\mu,\,\,\lambda d\lambda/\mu,\cdots,
\lambda^{g-1} d\lambda/\mu$ as a basis of holomorphic differentials on
$\Sigma_g$ but usually one takes linear combinations of these denoted
by $d\omega_j,\,\,1\leq j\leq g$, normalized via $\oint_{A_i}d\omega_j=
\delta_{ij}$, with period matrix defined via $\oint_{B_i}d\omega_j=
b_{ij}$.  The matrix $\Pi=(\Pi_{ij})$ is symmetric with $\Im\Pi>0$ and
it determines the curve. 
One will also want to consider another representation of hyperelliptic
curves of genus $g$ via
$\mu^2=\prod_0^{2g+1}(\lambda-\lambda_j)=P_{2g+2}(\lambda,\lambda_j)$
where $\infty$ is now not a branch point and there are two points
$\mu_{\pm}$ corresponding to $\lambda=\infty$.
\\[3mm]\indent
We extract now from \cite{ca} which provides some embellishments to
\cite{ff,kc}.
Variations of the Whitham-Bogoliubov averaging procedure have proved
to be of considerable interest and a very nice 
introduction to averaging for KdV appears in \cite{fd}.  Thus one
considers modulated waves based on a potential $u(\theta_i,\lambda_j)$ 
where the fast variables $x,\,t$ occur via $\theta_i= xU_i+tW_i$
and the slow variables occur in the moduli (here branch points) $\lambda_j
=\lambda_j(X,T)$ (for KdV one is dealing with hyperelliptic surfaces).  
The averaging procedure looks for example at
$<f(u)>\sim lim_{L\to\infty} (1/2L)\int_{-L}^Lf(u)\{dx\,\,or\,\,dt\}$ for
suitable functions $f(u)$, and, assuming incommensurable frequencies,
ergodicity implies $<f(u)>=(1/2\pi)^N\int_0^{2\pi}\cdots
\int_0^{2\pi}f(u(\theta_i,\lambda_j)\prod_1^{2N}d\theta_i$, where the
slow variables are fixed.  This is made more explicit below. 
One should perhaps
contrast this procedure explicitly with dispersionless limit techniques
which arise in the genus zero situation.  For KdV $u_t+(3/2)uu_x-
(1/4)u_{xxx}=0$ (based e.g. on $L^2=\partial^2-u,\,\,B=\partial^3-(3/2)
u\partial-(3/4)u_x,\,\,\partial_tL^2=[B,L^2]$) one writes $\epsilon t
=T,\,\,\epsilon x=X$ to obtain formally $u_T+(3/2)uu_X=0$ as $\epsilon
\to 0$ (dispersionless KdV or Euler equation).  The background dKdV
theory for Lax operators involves $L\psi=\lambda\psi$ with $\psi=
exp[(1/\epsilon)S(X,T)]$ so for $P=\partial_XS,\,\,\lambda=P^2-U(X,T)$
where one usually assumes $u(x,t)=U(X,T)+O(\epsilon)$.  This last step
is never really discussed adequately in work on (algebraic type) 
dispersionless theory and we make a brief comment here.  In question
is the behavior of $u(x,t)=u(X/\epsilon,T/\epsilon)$ as $x,t\to\infty$
(or $\epsilon\to 0$ with $X,T$ fixed) and there is a priori no modulated
wave or quasiperiodic situation to make the analysis easier.  A detailed
analysis of KdV $\to$ Euler for example is developed in classical work
of Lax, Levermore, and Venakides for example (see \cite{la} for
references and discussion) and a weak limiting procedure is justified
in many situations.  Similar analysis also applies to nonlinear Schr\"odinger
and Toda equations.  There is however no general theory here and the analysis
becomes deep and technical.  The moral is that the algebraic limiting
procedure is justified sometimes in realistic situations and we simply
assume it to be OK whenever it arises.  The averaging procedure for
quasiperiodic
modulated waves on the other hand seems to be quite generally applicable.
\\[3mm]\indent
Let us sketch now the averaging procedure following \cite{kc} with
clarifications as in \cite{ca,ff} (cf. also \cite{kg}).  Consider KP
in the form $3u_yy+\partial_x(4u_t-6uu_x+u_{xxx})=0$ via compatibility
$[\partial_y-L,\partial_t-A]=0$ where $L=\partial^2-u$ and $A=\partial^3
-(3/2)u\partial + w\,\,(u\to -u$ in (\ref{psi}).
We have then
$(\partial_y - L)\psi = 0$ with $(\partial_t -A)\psi = 0$ and for the 
adjoint or dual wave function 
$\psi^{*}$ one writes $\psi^{*}L = -\partial_y\psi^{*}$
with $\psi^{*}A =- \partial_t\psi^{*}$ where $\psi^{*}(f\partial^j)\equiv
(-\partial)^j(\psi^{*}f)$.  We can use formulas (cf. (\ref{psi}) and
(\ref{star}))
$\psi = e^{px+Ey+\Omega t}\cdot\phi(Ux+Vy
+Wt,P)$ and
$\psi^* = e^{-px-Ey-\Omega t}\cdot\phi^*
(-Ux-Vy-Wt,P)$
to isolate the quantities of interest in averaging
(here $p=p(P),\,\,E = E(P),\,\,\Omega =
\Omega(P),$ etc.)
We think here of a general Riemann surface $\Sigma_g$ with holomorphic
differentials $d\omega_k$ and quasi-momenta and quasi-energies
of the form $dp=d\Omega^1,\,\,dE=d\Omega^2,\,\,d\Omega=d\Omega^3,\cdots
\,\,(p=\int_{P_0}^Pd\Omega^1$ etc.) where the $d\Omega^j=d\Omega_j=
d(\lambda^j+O(\lambda^{-1}))$ are meromorphic differentials of the second
kind.  Following \cite{kc} we normalize now via $\Re\int_{A_j}d\Omega^k=
\Re\int_{B_j}d\Omega^k=0$.  Then write e.g. 
$U_k=(1/2\pi i)\oint_{A_k}dp$ and $U_{k+g}=-(1/2\pi i)\oint_{B_k}dp\,\,
(k=1,\cdots,g)$ with similar stipulations for $V_k\sim\oint d\Omega^2,\,\,
W_k\sim\oint d\Omega^3,$ etc.  This leads to real $2g$ period vectors
and evidently one could also normalize via $\oint_{A_m}d\Omega^k=0$ or
$\Im\oint_{A_m}d\Omega^k=\Im\oint_{B_m}d\Omega^k=0$.
It is then immediate that 
$(\psi^{*}L)\psi = \psi^{*}L\psi + \partial_x(\psi^{*}(L^1\psi)) +
\partial^2_x(\psi^{*}(L^2\psi)) + \cdots$
where e.g. $\psi^*(u\partial^j)=(-\partial)^j(\psi^*u)$.
More precisely, for $A=\sum_0^ka_i\partial^i$, one has $\psi^*A=\sum_0^k
(-\partial)^i(\psi^*a_i)$ and $(\psi^*A)\psi=\sum_0^k\partial^j(\psi^*
(A^j\psi))$ with $A^0=A,\,\,A^1=-\sum_1^kia_i\partial^{i-1},\,\,A^2=
\sum_2^k[i(i-1)/2]a_i\partial^{i-2},\cdots$.
For $L$ and $A$ as in KP above
$L^1 = -2\partial,\,\,L^2 = 1$ and 
$A^1 = -3\partial^2+(3/2)u,\,\,A^2 = 3\partial,\,\,A^3 =
-1$.
\\[3mm]\indent
Now for averaging we think of 
$u = u_0([1/\epsilon]S(X,Y,T)|I(X,Y,T)) + \epsilon u_1(x,y,t) +
\epsilon^2u_2(x,y,t) + \cdots$
($I\sim$ moduli and $S\sim$ action in 
some sense)
with 
$\partial_X S = U,\,\,\partial_Y
S = V,$ and $\partial_T S = W$.  We think of expanding
about $u_0$ with $\partial_x\to\partial_x + \epsilon\partial_X$.
This step will cover both $x$ and $X$ dependence for subsequent averaging.  
Then look at the compatibility condition $(\ddagger):\,\,\partial_tL
-\partial_y A + [L,A] = 0$.  We will want the term of first
order in $\epsilon$ upon writing e.g. $L = L_0 + \epsilon L_1 + \cdots$
and $A = A_0 + \epsilon A_1 + \cdots$ where $L_0,A_0$ are to depend
on the slow variables $X,Y,T$, and this gives
$\partial_t L_1 - \partial_y A_1 + [L_0,A_1] + [L_1,A_0] + F = 0;\,\,
F=\partial_TL - \partial_YA  -L^1\partial_XA + A^1\partial_XL$.
Thus $F$ is the first order term involving derivatives in the slow
variables.
Next one writes 
$\partial_t(\psi^*(L_1\psi)) -\partial_y(\psi^*(A_1\psi)) = 
\psi^*\{L_{1t}-A_{1y} 
+[L_1,A] +[L,A_1]\}\psi 
= \psi^*(\partial_tL_1-\partial_yA_1 + [L_0,A_1] + [L_1,A_0])\psi +
\partial_x(\cdots)$
and via ergodicity in $x,y$, or $t$ flows, averaging of derivatives in
$x,y$, or $t$ gives zero so one obtains
the Whitham equations in the form $<\psi^*F\psi> = 0$
(this represents the first order term in $\epsilon$ - the slow variables
are present in $L_0,\,\,A_0,\,\,\psi,$ and $\psi^*$).  
In order to spell
this out in \cite{ff} one imagines $X,Y$, or $T$ as a parameter $\xi$ and 
considers $L(\xi),\,\,A(\xi)$, etc. (in their perturbed form) with
$\psi(\xi) = e^{p(\xi)x+E(\xi)y+\Omega(\xi)t}\cdot
\phi(U(\xi)x + V(\xi)y 
+ W(\xi)t|I(\xi))$
and $\psi^* = exp(-px-Ey-\Omega t)\phi^*(-Ux-Vy-Wt|I)$ 
(no $\xi$ variation - i.e. assume $p,E,\Omega,U,V,W,I$ fixed).
Also note that
$x,y,t$ and $X,Y,T$ can be considered as independent variables.
Upon differentiating various expressions with respect to $x,\,y,\,t$
and $\xi$ and combining one obtains then
\be
d\Omega<\psi^*\psi> = -dp<\psi^*A^1\psi>;\,\,dE<\psi^*\psi> = 
-dp<\psi^*L^1\psi>
\label{BK}
\ee
and a version of the Whitham equations in the form
\be
p_T = \Omega_X;\,\,p_Y = E_X;\,\,E_T = \Omega_Y
\label{BM}
\ee
Note also from $\partial_XS=U,\,\,\partial_YS=V,$ and $\partial_TS=W$ one
has compatibility relations $\partial_YU=\partial_XV,\,\,\partial_TU=
\partial_XW,$ and $\partial_TV=\partial_YW$.
We mention also an equation obtained along the way which will be 
important in the analysis of \cite{kg}, namely
$$<\psi^*(L^1\partial_XA - A^1\partial_XL)\psi> = \partial_X\Omega
<\psi^*L^1\psi> - \partial_XE<\psi^*A^1\psi> +$$
\be
+ \partial_XW\cdot<\psi^*L^1\psi_{\theta}> - \partial_XV\cdot
<\psi^*A^1\psi_{\theta}>
\label{BG}
\ee
We feel that this derivation is important since it
exhibits again (as in \cite{fd} for the KdV situation)
the role of ``square eigenfunctions" (now in the form
$\psi^*\psi$) in dealing with averaging processes
(cf. \cite{ca,ce} for more on $\psi^*\psi$ and \cite{ca} for
details on averaging).
\\[3mm]\indent
Next, d'apr\`es \cite{ca}, we give some formulas for
differentials followed
by a few remarks on hyperelliptic situations.
The KP
flows can be written as $\partial_nu = K_n(u)$ where the $K_n$ are 
symmetries satisfying (in the notation of \cite{ce})
the linearized KP equation $\partial_3\beta = (1/4)\partial^3\beta
+ 3\partial(u\beta) + (3/4)\partial^{-1}\partial_2^2\beta = K'[\beta]$.
The conserved densities or gradients 
$\gamma$ satisfy the adjoint linearized KP
equation $\partial_3\gamma = (1/4)\partial^3\gamma + 3u\partial\gamma
+ (3/4)\partial^{-1}\partial_2^2\gamma$.  Then, replacing the square
eigenfunctions of KdV theory
by $\psi\psi^{*}$ one has e.g. $\psi\psi^{*} = \sum_0^
{\infty}s_n\lambda^{-n}$ where $s_n\sim\gamma_n$.  
Further $\partial_nu
= K_{n+1} = \partial s_{n+1} = \partial Res\,L^n = \partial\nabla\hat
{I}_{n+1}$ where $\nabla f\sim\delta f/\delta u$
($Res\,L^n = nH^1_{n-1}$ is generally used in the multipotential
theory).  We are thinking here in a single potential theory where all
potentials $u_i$ in $L = \partial + \sum_1^{\infty}u_{i+1}\partial^{-i}$
are expressed in terms of $u_2=u$ via operations with $\partial$ and
$\partial^{-1}$.  One uses here the Poisson bracket $\{f,g\} =
\int\int(\delta f/\delta u)\partial(\delta g/\delta u)dxdy$ (Gardner
bracket).   Thus 
one has $s_{n+1}\sim\gamma_{n+1}\sim\nabla\hat{I}_{n+1}$
as conserved gradients satisfying the adjoint linear KP equation
(**) $\partial_t\gamma = (1/4)\partial^3\gamma + 3u\partial\gamma
+(3/4)\partial^{-1}\partial^2_y\gamma$.  Following \cite{fd}
one linearizes around a fixed finite zone
solution $u$ in the adjoint linear KP equation and puts this $u$ into
the $\hat{I}_j$ etc.; then averaging over the $\theta_i$ 
variables is carried out as before.
Also from $\psi\psi^*=\sum_1^{\infty}(s_n/\lambda^n)$ (cf. \cite{ce})
we can write 
$<\psi\psi^*> = \sum_1^{\infty}(<s_n>/\lambda^n)$
Such a series is natural from asymptotic expansions
but when $\psi,\,\,\psi^*$ are written in terms of theta functions it
requires expansion of the theta functions in $1/\lambda$ (such
expansions are documented in \cite{da}, p. 49 for example).  It is now
natural to ask whether one can express $dp,\,\,dE,\,\,d\Omega$ from
(\ref{BK}) in more detail.  
Some order of magnitude considerations suggest
$dp\sim<\psi\psi^*>d\lambda$ near $\infty$.
To prove this we refer to \cite{ce} for
background and notation (cf. also \cite{cf,cg} for dispersionless
genus zero situations).  We write $\partial 
= L + \sum_1^{\infty}\sigma_j^1 L^{-j}$ which implies that
$(\partial\psi/\psi)=\lambda + \sum_1^{\infty} 
\sigma_j^1\lambda^{-j}$.  Then 
$(log\psi)_x=\lambda +\sum_1^{\infty}\sigma_j^1\lambda^{-j}$ and
$\overline{(log\psi)_x} \sim <(log\psi)_x>= p
= \lambda + \sum_1^{\infty} <\sigma_j^1>\lambda^{-j}$.  But $s_{n+1} =
-n\sigma_n^1 - \sum_1^{n-1}\partial_j\sigma_{n-j}^1$ and one 
obtains $<s_{n+1}> =
-n<\sigma_n^1> -\sum_1^{n-1}<\partial_j\sigma_{n-j}^1> = -n<\sigma_n^1>$. Thus
$dp = d\lambda - \sum_1^{\infty}j<\sigma_j^1>\lambda^{-j-1}d\lambda = d\lambda
+\sum_1^{\infty}<s_{j+1}>\lambda^{-j-1}d\lambda=<\psi\psi^*>$ since
$\psi\psi^*=\sum_0^{\infty}s_n\lambda^{-n}$. 
The normalization is built into this construction.  
It is clear that $dp=<\psi\psi^*>
d\lambda$ cannot hold globally since $<\psi\psi^*>$ should have poles at
$D+D^*$ and the correct global statement follows from \cite{kc,kg},
namely
$(dp/<\psi\psi^*>)=d\hat{\Omega}$
where $d\hat{\Omega}$ is the unique meromorphic differential with a double
pole at $\infty$ and zeros at $D+D^*$.  To see this note
that $dp/<\psi\psi^*>$ will have zeros at the poles of $\psi\psi^*$ (i.e.
at $D+D^*$) and a double pole at $\infty$; this characterizes $d\hat{\Omega}$.
\\[3mm]\indent 
In summary we see that
the quantity $\psi^*\psi$ is seen to determine the Whitham hierarchy
(\ref{BM}) and the differentials $dp,\,\,dE,\,\,d\Omega,$ etc. via
\be
\frac{dp}{<\psi\psi^*>}=-\frac{dE}{<\psi^*L^1\psi>}=-\frac{d\Omega}
{<\psi^*A^1\psi>}=d\hat{\Omega}
\label{BBBB}
\ee
\indent
In the hyperelliptic case with say $R(\Lambda)=\prod_1^{2g+1}(\Lambda
-\Lambda_j)$
one can explicitly write down the differentials
$dp,\,\,d\Omega,$ etc. as $dp=P(\Lambda)/\sqrt{R},\,\,P(\Lambda)=
\Lambda^g+\sum_1^ga_i\Lambda^{g-i}$ and $d\Omega=Q(\Lambda)/\sqrt{R},\,\,
Q(\Lambda)=6\Lambda^{g+1}+\sum_0^gb_j\Lambda^{g-j}$.  Now from (\ref{BM})
$(\clubsuit)\,\,
\partial_Tdp=\partial_Xd\Omega=(d\Omega/dp)\partial_Xdp$.  The moduli
here are the branch points $\Lambda_j$ which depend on the slow 
variables so writing out $(\clubsuit)$, multiplying by $(\Lambda
-\Lambda_i)^{3/2}$, and passing to limits $\Lambda\to\Lambda_i$ yields
the Whitham equations as equations for Riemann invariants
$\Lambda_k$ in the form $\partial_T\Lambda_k=v_k(\Lambda_1,\cdots,
\Lambda_{2g+1})\partial_X\Lambda_k$ where the characteristic velocities
$v_k=\left.(d\Omega/dp)\right|_{\Lambda=\Lambda_k}$ have an elementary
expression (cf. \cite{ca,fd}).
One can also use the integrals of motion ($\sim\hat{I}_n$ above) as
moduli (see remarks later).

\begin{center}\footnotesize
\section{ACTION AND PREPOTENTIAL}
\end{center}
\renewcommand{\theequation}{3.\arabic{equation}}\setcounter{equation}{0}

\normalsize

The viewpoint of \cite{ib} for example is now to look at SW theory in
terms of a renormalization group map $\{G,\tau,m,h_i\}\to\{a_i(h),{\cal F}
(a_i)\}$ where $G\sim$ gauge group, $\tau\sim$ bare coupling constant,
$m\sim$ mass scale, $h_i\sim$ symmetry breaking vev's, $a_i(h)\sim$
background ``fields", and ${\cal F}\sim$ prepotential (with $a_i=
\partial{\cal F}/\partial a_i$, etc.).  This map is then decomposed as
$\{G,\tau,m,h\}\to\{\Sigma,d{\cal S}_{min}\}\to\{a_i(h),{\cal F}(a_i)\}$
where $\Sigma$ is a Riemann surface and $d{\cal S}_{min}$ is a meromorphic
one form on $\Sigma$ such that $\partial d{\cal S}/\partial h_i$ is
holomorphic.  Since the genus of $\Sigma$ may be larger than $r_G=rank(G)$
one must make some adjustments for which we refer to \cite{ib}.  Roughly
$d{\cal S}_{min}\sim d\lambda_{SW}$ (modulo a few nonzero 
times - cf. \cite{cz,gd,ib}) and 
$h\sim$ moduli as will be indicated.  The first map to $\{\Sigma,
d{\cal S}_{min}\}$ involves quite general ideas and can be viewed via a
spectral cover of some base space $E(\tau)$ and an integrable system
determined by a (matrix) Lax operator based on $E(\tau)$ (generally
$E(\tau)\sim{\bf P}^1$ or an elliptic curve).  Thus one looks at
$\Sigma:\,\,det(t-L(z))=0$ which defines $\Sigma$ as a ramified cover
of $E$ and the integrals of motion (or Casimirs) for the corresponding
integrable system will be identified with moduli $h_i$.  This is in keeping
with the spirit of averaging in Section 2.  In this context $d\lambda_{SW}
\sim td\omega$ where $d\omega$ is (the) normalized holomorphic differential
on $E(\tau)$ and this will be clarified below (cf. \cite{cz,dg,ib,kg}).
Regarding the second map the action integrals $a_i=S_{A_i}=\oint_{A_i}d\lambda_
{SW}$ and $a_i^D=S_{B_i}=\oint_{B_i}d\lambda_{SW}$ for $(A_i,B_i)$ a 
canonical homology basis will play a fundamental role.  To assist in the
determination of $d\lambda_{SW}$ we recall that the monodromy for
period integrals $\oint_Cd\omega\,\,(d\omega_i$ a basis of
normalized holomorphic differentials) is often obtained from the physics
and one will have PF equations for period integrals, etc.  Thus it will be
sufficient to have $\partial d\lambda_{SW}/\partial h_k=\sum_1^g\alpha_{kj}
d\omega_j$ so e.g. $\partial a_i/\partial h_k=\oint_{A_i}\sum\alpha_{kj}
d\omega_j=\sum\alpha_{kj}\delta_{ji}=\alpha_{ki}$ and $\partial a_i^D/
\partial h_k=\oint_{B_i}\sum\alpha_{kj}d\omega_j=\sum\alpha_{kj}b_{ji}$
where $(b_{ij})$ is the period matrix for $\Sigma$ (cf. also \cite{kg} and
remarks below).  For hyperelliptic $\Sigma$ one often takes $\partial
d\lambda_{SW}/\partial v_k=d\hat{\omega}_k=(\Lambda^kd\Lambda/y)\sim$
unnormalized holomorphic differentials ($y=\sqrt{R},\,\,R=\prod_1^{2g+2}
(\Lambda-\Lambda_i),\,\,k=0,\cdots,g-1$) where $v_k\,\,(1\leq k\leq g)$
are suitable ``flat" moduli (see e.g. \cite{ea,kn,nb} for good illustrations
and calculations).
\\[3mm]\indent
Now from \cite{ib} (cf. also \cite{cd}) one works from a general action
$d{\cal S}$ and builds in times $T_n$ from the beginning.
For convenience the development here is in terms
of hyperelliptic spectral curves $\Sigma$ which arise automatically
from periodic Toda hierarchies (cf. \cite{be,cd,da}); physically the
framework is appropriate for $SU(n)\,\,N=2$ susy YM theories with matter
for example (and others).  Then the $T_n$ 
correspond to quasiclassical (slow) times and the prepotential ${\cal F}$
will correspond to $F=$ logarithm of the tau function for the dispersionless
Toda hierarchy.  For convenience the development here is in terms
of hyperelliptic spectral curves $\Sigma$ which arise automatically
from periodic Toda hierarchies (cf. \cite{be,cd,da}).
Thus let $d\omega_i$
be canonical holomorphic differentials on $\Sigma_g=\Sigma_g(h_k)$ where
$i=1,\cdots,g$ and $k=1,\cdots,K$.  The $h_k$ here are to represent the
moduli (e.g. they could correspond to  
branch points $\Lambda_k$ for hyperelliptic situations).
One has then canonical normalizations $\oint_{A_i}d\omega_j=
\delta_{ij}$ and $\oint_{B_i}d\omega_j=b_{ij}(h_k)$.
From the point of view of Yang-Mills (YM) theory 
the spectral curve $\Sigma_g$ is determined by the gauge
group $G$ with $h_k\sim (1/k)<Tr\phi^k>$ for a suitable scalar field
in the adjoint representation of $G$ which breaks the original gauge
symmetry down to $U(1)^{r_G}$ (cf. here \cite{bc,dg,sb} for details).
From the point of view of integrability the $(1/k)<Tr\phi^k>$ correspond
to Hamiltonians (integrals of motion of the integrable system) and one is
back to Whitham dynamics.  Now in \cite{ib} one begins with a meromorphic
differential $dS$ on $\Sigma_g$ defined (partly) by the requirement
$(\bullet)\,\,
(\partial dS/\partial h_k)\simeq\sum_1^g\sigma_{ki}^{dS}d\omega_i
\simeq$ holomorphic differential,
where everything depends on the $h_k$.  Existence of solutions depends
on the number $K$ of moduli and one will assume here that $K=g$ 
(one expects $K>g$ generically).
We will anticipate hyperelliptic curves here 
(e.g. Toda situations) and assume $\Sigma_g$ to be given by 
$\lambda^2=\prod_1^{2g+2}(\Lambda-\Lambda_i)=R_g(\Lambda)$
for example.
Let us pick two points $\lambda_{\pm}\in\Sigma_g$ and use $\lambda$ for a
complex coordinate near $\lambda_{+}$ or $\lambda_{-}$ ($\lambda$ is  
correct here, not $\Lambda$).  One defines
$d\hat{\Omega}_n$ as a solution to $(\bullet)$ satisfying in addition
(1) $d\hat{\Omega}_{\pm n}\,\,(n\geq 1)$ has a pole of order $n+1$ at
$\lambda_{\pm}$ and no other poles with $d\hat{\Omega}_{\pm n}(\lambda)=
\pm[(\lambda-\lambda_{\pm})^{-n-1}+o(1)]d\lambda$ and (2) $d\hat{\Omega}_0$
has simple poles at both $\lambda_{\pm}$ with residues $\pm1$.
Such $d\hat{\Omega}_n$ exist for $K=g$ for 
example and one has then $(\bullet\bullet)\,\,
(\partial d\hat{\Omega}_n)/\partial h_k)\simeq\sum_1^g\sigma_{ki}^n
d\omega_i$.  On the other hand differentials
$d\Omega_n$, satisfying (1) and (2),
but not necessarily $(\bullet\bullet)$, arise in Whitham theory with
normalization
$\oint_{A_i}d\Omega_n=0$.
Note that the conditions (1) and (2) plus normalization define the
$d\Omega_n$ and then $d\hat{\Omega}_n\simeq d\Omega_n+\sum_1^gc_i^nd\omega_i$
must hold with $c_i^n=\oint_{A_i}d\hat{\Omega}_n$.  
One can write then $\kappa_j^n=
\oint_{B_j}d\Omega_n$ and refer to \cite{nb} for
the tau function of Toda theory.
Then one could use $\psi^*\psi$ as in Section 2 for averaging or work
directly from an action expression (a nice exposition of Toda averaging
is in \cite{be}) and  
end up with an action term $d{\cal S}(h_k,T_i)=
\sum_1^ga_id\omega_i+\sum_{-\infty}^{\infty}T_nd\Omega_n$ where the $h_k$
are moduli and the $T_n$ are slow variables.  A priori $a_i,\,\,d\omega_i,$
and $d\Omega_n$ depend on $h_k$ and $T_n$ and further analysis is
mandatory.
\\[3mm]\indent
Thus to be more systematic in developing a
quasiclassical tau function whose logarithm corresponds
to the Seiberg-Witten (SW) prepotential one takes a solution
$d{\cal S}(T_n|h_k)$ of $(\bullet)$ 
(presumed to exist without loss of generality since we have indicated
how such creatures could arise from period integrals and monodromy)
such that (*) $(\partial d{\cal S}/
\partial T_n)\simeq d\Omega_n$, which would make $d{\cal S}$ a kind of
generating function for all solutions of $(\bullet)$.  Thus write
at the quasiclassical or averaged level 
$a_j=\oint_{A_j}d{\cal S};\,\,a_j^D=\oint_{B_j}d{\cal S}$
and stipulate that
$(\partial a_j/\partial T_n)=0$ and $(\partial a_j^D/
\partial T_n)=\oint_{B_j}d\Omega_n=\kappa_j^n(h)$.
Now expand $d{\cal S}$ as $(\dagger)\,\,
d{\cal S}\simeq\sum_{-\infty}^{\infty}u_m(T)d\hat{\Omega}_m(h)$
with coefficients $u_m(T)$ independent of $h$.  Putting $(\dagger)$ into
(*) we get
\be
\frac{\partial d{\cal S}}{\partial T_n}\simeq\sum_{-\infty}^{\infty}\left(
\frac{\partial u_m}{\partial T_n}d\hat{\Omega}_m+u_m\frac{\partial
d\hat{\Omega}_m}{\partial h_k}\frac{\partial h_k}{\partial T_n}\right)\simeq
\label{CO}
\ee
$$\simeq\sum_{-\infty}^{\infty}\left(\frac{\partial u_m}{\partial T_n}
d\hat{\Omega}_m+u_m\sum_k\frac{\partial h_k}{\partial T_n}\sum_1^g
\sigma_{ki}^m(h)d\omega_i\right)$$
We will write now $\partial_n\equiv\partial/\partial T_n$ and then 
upon comparing
$\partial_nd{\cal S}\simeq d\Omega_n\simeq d\hat{\Omega}_n-\sum_1^g
c_i^nd\omega_i$ with (\ref{CO}) one gets $(\clubsuit\clubsuit)\,\,
\partial_nu_m=\delta_{mn}\Rightarrow u_m(T)=T_m$ along with the formula
$\sum_k\partial_nh_k\left(\sum_{-\infty}^{\infty}T_m\sigma_{ki}^m\right)
\equiv\sum_k\partial_nh_k\sigma_{ki}=-c_i^n$.
The last equation represents a version of 
the Whitham dynamics for $h_k$ ($\sigma_{ki}$ is defined
as indicated).  
In particular the moduli $h_k$ are necessarily $T$ dependent.
We note that normally one thinks of Whitham equations as equations for
$\partial_n\Omega_k$ as in (\ref{BM}) or as equations for branch points
as indicated after (\ref{BBBB})
(here $\partial_n\sim\partial/\partial T_n$ for slow
variables $T_n$).  The branch points are moduli in these hyperelliptic
situations so in a general sense one can think of any nonlinear first
order PDE in the moduli (e.g. $h_k$) as Whitham equations.  In the present
situation one has in fact expressions $(\partial d\hat{\Omega}_n/\partial
h_k)=\sum\sigma_{ki}^nd\omega_i$ with $\partial_nd{\cal S}=d\hat{\Omega}_n-
\sum c_i^nd\omega_i$ whose combination leads to $(\clubsuit\clubsuit)$,
i.e. to 
$\sum\partial_nh_k\sigma_{ki}=-c_i^n$.  At this point no restrictions
on the number of moduli have been made (but see below).
We refer also to \cite{bx} for another viewpoint on Whitham theory.
\\[3mm]\indent
Now $(\dagger)$ and $(\clubsuit\clubsuit)$ imply (using 
$(\spadesuit\spadesuit)\,\,d\hat{\Omega}_n\simeq d\Omega_n+\sum_1^g
c_i^n(h)d\omega_i;\,\,c_i^n(h)=\oint_{A_i}d\hat{\Omega}_n)$ that one
has
$d{\cal S}\simeq\sum_{-\infty}^{\infty}T_md\hat{\Omega}_m(h|T)\simeq
\sum_{-\infty}^{\infty}\left(T_md\Omega_m+T_m\sum_1^gc_i^md\omega_i\right)$
Integrating this along $A_j$ cycles and using 
$(\spadesuit\spadesuit)$ again we get
$a_j=\oint_{A_j}d{\cal S}=\sum_{-\infty}^{\infty}T_mc_j^m=\sum_{-\infty}^
{\infty}T_m\oint_{A_j}d\hat{\Omega}_m$
leading to the desired expression
$d{\cal S}\simeq\sum_1^ga_id\omega_i+\sum_{-\infty}^{\infty}T_md\Omega_m$.
Note that $a_i=a_i(h(T))$ means that the $a_i$ seem to be $T$ dependent
but the $T$ dependence via $h_k$ actually cancels out since 
$\partial a_j/\partial T_n=0$ has been stipulated.
This means that one can add the $a_i$ as additional independent
variables to the set of slow times $T_n$ and write $d{\cal S}=
d{\cal S}(a_i,T_n)$.  Upon doing this we should add an equation for
$\partial d{\cal S}/\partial a_i$ to (*).  This can not be postulated
arbitrarily since $d{\cal S}$ is already defined and hence such equations
must be derived.  To do this we assume now explicitly $K=g$ so 
that $(\sigma_{ki})$ in $(\clubsuit\clubsuit)$ will be a square
matrix.  Then some calculation yields $(\left.\partial h_k/\partial a_j
\right|_{T=c})=(\sigma_{jk})^{-1}(h)$ from which one can conclude that
$d{\cal S}=\sum_{-\infty}^{\infty}T_md\hat{\Omega}_m=\sum_1^ga_id\omega_i
+\sum_{-\infty}^{\infty}T_nd\Omega_n$ with $\partial d{\cal S}/\partial
a_i=d\omega_i$ and $\partial_nd{\cal S}=d\Omega_n$.
We can also write now $h_k(a_i,T_n)$ once the $a_i$ have been established
as independent variables.
Next one looks at the B-periods of $d{\cal S}$ in the form
\be
a_j^D=\oint_{B_j}d{\cal S}=\oint_{B_j}\left(\sum_1^ga_id\omega_i+\sum_
{-\infty}^{\infty}T_nd\Omega_n\right)=\sum_1^ga_ib_{ij}(h)+
\sum_{-\infty}^{\infty}T_n\kappa_j^n(h)
\label{DA}
\ee
Then evaluating 
$\partial a_j^D/\partial a_i=\oint_{B_j}(\partial d{\cal S}/
\partial a_i)=\oint_{B_j}d\omega_i=b_{ij}(h)$
we see from the symmetry $b_{ij}=b_{ji}$ that there is some
function ${\cal F}$ such that 
$a_j^D=(\partial{\cal F}/\partial a_j)$.
This ${\cal F}(a_i,T_n)$ is called a logarithm of a quasiclassical tau
function or a prepotential.  The arguments are essentially the moduli
of solutions to the original Toda like or low energy YM dynamical
system with $a_i(h_k)$ parametrizing the moduli of the curve $\Sigma_g$ and
$T_n$ moduli for coordinate systems in the vicinity of punctures.  It turns
out that the point $T_n=0$ for all $n$ is usually singular (note
$\sigma_{ki}=\sum_{-\infty}^{\infty}T_m\sigma_{ki}^m$ vanishes when all
$T_n=0$ so $\sigma_{ik}^{-1}$ does not exist). 
One can give explicit formulas for many quantities in terms of 
contour integrals which we omit here (cf. \cite{ib}) but we do note
that ${\cal F}$ is a homogeneous function of degree two, namely
${\cal F}=\frac{1}{2}\left(\sum_1^ga_i(\partial{\cal F}/\partial a_i)
+\sum_{-\infty}^{\infty}T_n\partial_n{\cal F}\right)$.

\begin{center}\footnotesize
\section{SYMPLECTIC GEOMETRY}
\end{center}
\renewcommand{\theequation}{4.\arabic{equation}}\setcounter{equation}{0}

\normalsize

The study of TFT and LG theories in connection with integrable systems
and Whitham dynamics has an extensive literature (we mention here only
\cite{ab,ca,cd,cf,cg,db,gx,kd,tb}).  There are many visible
connections between $F$ in the quasiclassical tau function $\tau=
exp(F)$ for KP/Toda theories and the prepotential ${\cal F}$.  Further,
${\cal F}$ and the LG (super)potential $W$ for example are also related
to the Hodge-K\"ahler special geometry in the $N=2$ wonderland (cf.
\cite{cd,fh,gx}).  Recently in \cite{bg} it was shown that the theory
of the prepotential for $N=2$ susy YM leads to WDVV (Witten-Dijkgraaf-
Verlinde-Verlinde) equations as in TFT (cf. \cite{db}).  Also 
in \cite{fx} one uses duality theory as in 
SW theory to produce a Legendre transform relating $x,\,\,{\cal F},$ and
the probability density $|\psi|^2$ for the Schr\"odinger equation; this
is suggested as a way to quantize geometry.  Thus there are many
directions and connections.
\\[3mm]\indent
In this section we discuss briefly some constructions from \cite{kg}
which were in part spelled out more completely in \cite{cd}.
The paper \cite{kg} gives a particularly appropriate development of
the theory of integrable systems in connection with $N=2$ susy YM.
The role of $d\lambda_{SW}$ as an action-period creature is emphasized
and for hyperelliptic Toda systems of the type arising in $N=2$ susy
YM the theory seems quite complete.  For systems
of Calogero-Moser type with underlying elliptic curve there is much
less information. In any event one 
outstanding feature of this work is to show how much can be based on
the BA functions $\psi$ and $\psi^*$.
One begins with a universal configuration space which is a moduli space
of Riemann surfaces with $N$ punctures $P_{\alpha}$
(better thought of as marked points)
and two Abelian integrals $E$ and $Q$ with poles of order at most
$n=(n_{\alpha})$ and $m=(m_{\alpha})$ at the $P_{\alpha}$.  One defines
an $n_{\alpha}$-jet $[z_{\alpha}]_{n_{\alpha}}$ of coordinates near a
puncture $P_{\alpha}$ to be an equivalence class of coordinates $z_{\alpha}$
with $z_{\alpha}\equiv z'_{\alpha}$ if $z'_{\alpha}=z_{\alpha}+O(z_{\alpha}^
{n_{\alpha}+1})$; this space of jets has dimension equal to $n_{\alpha}$.
For $[z]_n$ a jet near $P_1$ we define an Abelian integral $Q$ as a pair
$(dQ,c_Q)$ where $dQ$ is a meromorphic differential on the surface $\Sigma$ and
$Q=\sum_{-m}^{\infty}c_kz^k+c_Q+R^Qlog(z)$ if 
$dQ=d(\sum_{-m}^{\infty}c_kz^k)+R^Q(dz/z)$.  By integrating $dQ$ along
paths one extends the Abelian integral $Q$ holomorphically to a neighborhood
of any point in $\Sigma/\{P_1,\cdots,P_N\}$.  The analytic continuation
will depend in general on the path. 
Let us concentrate on $N=1$, with $(n,m)=(n,1)$ fixing the singularities
for $E$ and $Q$, and take $dE\sim
d(z^{-n}+O(z))dz \sim d\Omega_n$ with $dQ\sim d(z^{-1}+O(z))\sim dp$
(necessarily $R_1^E=R_1^Q=0$).  For comparison purposes below we will take
$n=3$ later as well.  The universal configuration space is defined as
${\cal M}_g(n,1)=\{\Sigma,P_1,[z]_n,E,Q\}$ where $\Sigma$ is a smooth
Riemann surface of genus $g$; 
there are then $3g-3$ parameters for the Riemann surface,
$n$ for the jet $[z]$ or order $n$, and $g+1$ each for $E$ and $Q$, giving
$5g+n$ parameters (note e.g. $E$ has one parameter for $c_E$ and $g$ for the
holomorphic differentials which can be added without modifying the
singular expansion.  For local coordinates one chooses now
$5g+n$ objects ($P\sim P_1,\,\,z\sim z_1$)
\be
T_k=\frac{1}{k}Res_P(z^kQdE),\,\,(k=1,\cdots,n);\,\,
\tau_{A_i,E}=\oint_{A_i}dE;
\label{GC}
\ee
$$\tau_{B_i,E}=\oint_{B_i}dE;\,\,
\tau_{A_i,Q}=\oint_{A_i}dQ;\,\,
\tau_{B_i,Q}=\oint_{B_i}dQ;\,\,a_i=\oint_{A_i}QdE$$
where $i=1,\cdots,g$ and it is $QdE=pdE$ which will play the role of
$d\lambda_{SW}$ here.
\\[3mm]\indent
Now let ${\cal D}$ be the open set in ${\cal M}_g(n,1)$ where the zero
divisors of $dE$ and $dQ$ do not intersect (i.e. the sets $\{\gamma;\,\,
dE(\gamma)=0\}$ and $\{\gamma;\,\,dQ(\gamma)=0\}$ do not intersect).
It is proved that near each point in ${\cal D}$ the $5g+n$ functions
$T_k,\,\,\tau_{A_i,E},\,\,\tau_{B_i,E},\,\,
\tau_{A_i,Q},\,\,\tau_{B_i,Q},\,\,a_i$ have linearly independent differentials,
and thus define a local holomorphic coordinate system.  Further the joint
level sets of these functions (omitting the $a_i$) define a smooth
g-dimensional foliation of ${\cal D}$, independent of the choices made in
defining the coordinates themselves.
Now ${\cal M}_g={\cal M}_g(n,1)$ can be taken as a base space for two
fibrations ${\cal N}^g$ and ${\cal N};\,\,{\cal N}^g$ has fiber
$S^g(\Sigma)\simeq J(\Sigma)=$ Jacobian variety (via the Abel map
$(\gamma_1,\cdots,\gamma_g)\to \sum_{i=1}^g\int_P^{\gamma_i}d\omega_j)$
and ${\cal N}$ has fiber $\Sigma$ (all over a point $(\Sigma,P,[z]_n,
E,Q)\in{\cal M}_g$).  We consider only leaves ${\cal M}$ of the foliation
of ${\cal D}$ indicated above and look at the fibration ${\cal N}$ or
${\cal N}^g$ over the base ${\cal M}$.
One wants to define a symplectic form $\omega_{{\cal M}}$
on ${\cal N}^g$.  First, although $E$ and $Q$ are multivalued on
the universal fibration their differentials are well defined on ${\cal N}$.
The idea here is that $E$ and $Q$ are well defined near $P_1$ and their
analytic continuations by different paths can only change by multiples
of their residues or periods along closed cycles.  But on a leaf of the
foliation the ambiguities remain constant and disappear upon differentiation.
Hence one has differentials $\delta E$ and $\delta Q$ on the fibrations
which reduce to $dE$ and $dQ$ acting on vectors tangent to the fiber. 
One can trivialize the fibration ${\cal N}$ with the variables
$a_1,\cdots,a_g$ along the leaf ${\cal M}$ and e.g. $E$ along the
fiber.  Then $dQ$ coincides with $(dQ/dE)dE$ where $dE$ is viewed as one
of the elements of the basis of one forms for ${\cal N}$ 
and the full differential is $\delta Q=dQ+\sum_1^g(\partial Q/
\partial a_i)da_i\equiv dQ+\delta^EQ$
(also one takes $c_E=0$). 
\\[3mm]\indent
Now if one considers the full differential $\delta(QdE)$ on ${\cal N}$
it is readily seen that it is well defined despite the multivaluedness
of $Q$.  In fact the partial derivatives $\partial_{a_i}(QdE)$ along
the base ${\cal M}$ are holomorphic since the singular parts of the 
differentials as well as the ambiguities are all fixed.  In particular
$(\partial/\partial a_i)(QdE)=d\omega_i$
where $d\omega_i$ is a basis of normalized holomorphic differentials
$\oint_{A_i}d\omega_j=\delta_{ij}$ with $\oint_{B_i}d\omega_j=b_{ij}$.
To see this note that 
it is implicit in (\ref{GC}) since by definition of the $a_i,\,\,
(\partial a_i/\partial a_j)=\delta_{ij}=\oint_{A_i}(\partial (QdE)/
\partial a_j)$ which implies $(\partial (QdE)/\partial a_j)=d\omega_j$.
Formally then one defines $\omega_{{\cal M}}$ on ${\cal N}^g$ via
\be
\omega_{{\cal M}}=\delta\left(\sum_1^gQ(\gamma_i)dE(\gamma_i)\right)=
\sum_1^g\delta Q(\gamma_i)\wedge dE(\gamma_i)=\sum_1^gda_i\wedge d\omega_i
\label{GF}
\ee
The first expression seems formally reasonable on ${\cal N}^g$ and the last
appears to be a calculation of the form
$\omega_{{\cal M}}=\delta(QdE)=\sum_1^gda_i\wedge [\partial(QdE)/\partial
a_i]=\sum_1^gda_i\wedge d\omega_i$.
Now go to KP for illustration and background.  One can work with 
a (nonspecial) divisor
$(\gamma_1,\cdots,\gamma_g)$ giving rise to quasiperiodic
functions of $t=(t_n)\,\,1\leq n<\infty,$ of the form $u_{i,n}
\,\,1\leq i\leq n-2,\,\,2\leq n<\infty$, which arise as solutions of an
integrable hierarchy.  The BA function is defined as a meromorphic function
away from $P$ with simple poles at the $\gamma_i\,\,(1\leq i\leq g)$
and an essential singularity at $P$ of the form
$\psi(t,z)=exp\left(\sum_1^{\infty}t_nz^{-n}\right)
\left(1+\sum_1^{\infty}\xi_i(t)z^i\right)$.
There is a Lax operator (in general one for each puncture)
$L_n=\partial^n+
\sum_0^{n-2}u_{i,n}\partial^i$
with $\left(\partial/\partial t_n-L_n\right)
\psi(t,z)=0\,\,(\partial=\partial_x,\,\,x=t_1)$.
Thus there is a map $\{\Sigma,P_,z,
\gamma_1,\cdots,\gamma_g\}\to \{u_{i,n}(t)\}$.  
An explicit form for the BA function for KP is given in \cite{kg}
but it involves normalizations $\Re\oint_Cd\Omega_n=0$ for any cycle
$C$ and is more complicated in appearance that our previous expression
in Section 2 for example.  Similarly the dual BA function $\psi^*$ is 
defined as before (and denoted by $\psi^{\dagger}$ in \cite{kg}).
We note that for the Toda lattice one takes $N=2$ punctures.
\\[3mm]\indent
An element
in ${\cal N}^g(n,1)$ gives rise to a datum in a space 
$\hat{{\cal N}}^g$ via
$\Xi:\,\,(\Sigma,P,[z]_n,E,Q,\gamma_1,\cdots,\gamma_g)\to (\Sigma,
P,z,\gamma_1,\cdots,\gamma_g)\to \{\left.u_{i,n}(t)\right|_{i=1}^{n-2}\}
\in \hat{{\cal N}}^g$.
Take a real leaf ${\cal M}$ (i.e. $\Re\oint_CdE=\Re\oint_CdQ=0$
for all cycles $C$ on $\Gamma$) 
and write still
$dQ\sim d\Omega_1=dp$ and $dE\sim d\Omega_n$
(with real normalizations).
We also take $t_1\sim x$.  One wants now
to express $\omega_{{\cal M}}$ in terms of forms on the space of functions
$\{u_{i,n}(t)\}$.  First the $u_{i,n}(t)$ can be written in terms of the 
asymptotic BA coefficients
$\xi_i$ and one knows that
$(\partial_x\psi/\psi)=z^{-1}+\sum_1^{\infty}h_sz^s$ and $
(\partial_x\psi^*/\psi^*)=-z^{-1}+\sum_1^{\infty}h^*_sz^s$
(cf. \cite{ce} and arguments before (\ref{BBBB}) where e.g.
$\partial = L+\sum_1^{\infty}\sigma_j^1L^{-j}$  which 
implies e.g. $(\partial\psi/\psi)=\lambda+\sum_1^{\infty}\sigma_j^1
\lambda^{-j}$ for $\lambda\sim z^{-1})$.  In any event the first $n-1$
coefficients $h_s,\,\,h^*_s,$ or $\sigma_j^1$ are differential polynomials
in the $u_{i,n}$ (initial data $\left.\xi_s(t)\right|_{x=0}=\phi_s(t_2,\cdots)$
determine $\xi_s$ for $s\leq n-1$).  Writing $H_s=<h_s>$
one gets now $p=z^{-1}+\sum_1^{\infty}
H_sz^s$ (cf. \cite{ca}).  We note that one uses $<\,\,\,>_x$ and $<\,\,\,>_{xy}$
averaging at various places in \cite{kg} but generically this should
correspond to ergodic averaging.  
\\[3mm]\indent
A result in \cite{kg} now asserts that
for ${\cal N}^g$ the Jacobian
bundle over a real leaf ${\cal M}$ of the moduli space ${\cal M}_g(n,1)$
the symplectic form $\omega_{{\cal M}}$ can be written as
\be
\omega_{{\cal M}}=-Res_P\frac{<\delta\psi^*\wedge\delta L\psi>}{<\psi^*\psi>}dp
=n\sum_1^{n-2}<\delta h_s\wedge\int^x\delta^*h^*_{n-s}>
\label{GK}
\ee
where the $h_s,\,\,h^*_s$ are differential polynomials as above, and the
differential forms $\delta h_s$ and $\delta^*h^*_s$ are defined via
$\delta h_s=\sum_{i=0}^{n-2}\delta u_{i,n}\sum_{\ell}(-\partial)^{\ell}
(\partial h_s/\partial u_{i,n}^{(\ell)})$ and  
$\delta^*h^*_s=\sum_{i=0}^{n-2}\delta u_{i,n}^{(\ell)}\sum
(\partial h^*_s/\partial u_{i,n}^{(\ell)})$.
\\[3mm]\indent
This is a fascinating result but the proof in \cite{kg} requires some
embellishment.  First one must come to terms with an expression
$(\clubsuit\spadesuit)\,\,
\delta E=\delta p(dE/dp)+(<\psi^*\delta L_n\psi>/<\psi^*\psi>)$
where $\delta L_n=\sum_0^{n-2}\delta u_{i,n}\partial^i$
(see below) and then it will follow from (\ref{GF}) (with $\delta p\sim
\delta Q$) that $(\clubsuit\bullet)\,\,\omega_{{\cal M}}=-\sum_1^g(<\psi^*
\delta L_n\psi>/<\psi^*\psi>)(\gamma_s)\wedge \delta p(\gamma_s)$.
The formula $(\clubsuit\spadesuit)$ is asserted to come from \cite{kc}
but to see this one has to interpret $\delta u_{i,n}$ as arising from
$\epsilon\partial_XU_{i,n}(X,T)$ in the first order term.  Indeed we
can look at (\ref{BG}) with the last two terms absent on the leaves of
our foliation and written generically for $L_n\sim \Omega_n$ as
$(\bullet\spadesuit)\,\,<\psi^*(L_n^1\partial_XL_k-L_k^1\partial_XL_n)
\psi>=\partial_X\Omega_k<\psi^*L_n^1\psi>-\partial_X\Omega_n
<\psi^*L_k^1\psi>$.  Then for $L_n=L_3\sim\Omega_3$ and $L_k=L_1\sim
p$ one has $L_1=\partial,\,\,\delta L_1\sim\partial_XL_1=0,\,\,L_1^1=1$
and $-<\psi^*\partial_XL_3\psi>=\partial_Xp<\psi^*L_3^1\psi>-\partial_X
\Omega_3<\psi^*\psi>$.  Written in terms of $\delta p\sim\epsilon
\partial_Xp,\,\,\delta L_3\sim\epsilon\partial_XL_3$, etc. one obtains
$\delta\Omega_3=\delta p(<\psi^*L_3^1\psi>/<\psi^*\psi>)+(<\psi^*
\delta L_3\psi>/<\psi^*\psi>)$ while from (\ref{BBBB}) (with $L_3^1\sim
A^1$) we have $(<\psi^*L_3^1\psi>/<\psi^*\psi>)=(d\Omega_3/dp)$, 
or generically $(d\Omega_n/dp)=(<\psi^*L_n^1\psi>/
<\psi^*\psi>)$, and 
$(\clubsuit\spadesuit)$ follows.  
The formula $(\bullet\spadesuit)$ can also be used
for a more general theorem in \cite{kg}.  Thus $(\clubsuit\bullet)$ holds and
writing $\gamma_s(t)$ for the zeros of $\psi$ away from $P$ (corresponding
to the fixed poles $\gamma_s$ for $t=0$) one can pick $t_1,\cdots,t_g$ 
(generically) as times for which the flows $\gamma_s(t)$ are independent
and use them as coordinates on $S^g(\Sigma)$.  These can be transferred to
the system of coordinates $f(\gamma_1),\cdots,f(\gamma_g))$ for $f$ an
Abelian integral on $\Sigma$ via 
$(\partial/\partial t_i)f(\gamma(t))=Res_{\gamma(t)}
[(\partial/\partial t_i)\psi(t,z)/\psi(t,z)]df$.  In the present
case, writing $(\delta_t\psi/\psi)=\sum_1^g(\partial_j\psi/\psi)dt_j$ and 
using $\sum Res_{\gamma_s}=-Res_P$ one obtains 
$\omega_{{\cal M}}=Res_P\left[(<\psi^*\delta L_n\psi>/
<\psi^*\psi>)\wedge
\left.(\delta_t\psi/\psi)\right|_{t=0}\right]dp$.
It is not unnatural to see here an apparent
change in the number of parameters.  Now note that $\psi^*\delta L_n\psi=
\psi^*\sum_0^{n-2}\delta u_{i,n}\partial^i\psi
=\sum_0^{n-2}\delta u_{i,n}\psi^*\partial^i\psi$ and an easy
calculation gives $(\partial^j\psi/\psi)=z^{-j}(1+\sum_1^{\infty}
c_{jp}z^p)$; this implies 
$(\psi^*(\delta L_n\psi)/<\psi^*\psi>)dp=
-\sum_1^{\infty}\delta J_sz^{-n-1+s}dz$ for coefficients
$\delta J_s\sim$ linear combinations of the $\delta u_{i,n}$ (recall
$(dp/<\psi^*\psi>)=O(z^{-2})$).  Averaging now gives 
$(<\psi^*(\delta L_n\psi>/<\psi^*\psi>)dp=
-\sum_1^{\infty}<\delta J_s>z^{-n-1+s}dz$ so 
$(\clubsuit\spadesuit)$ implies $\delta H_s=<\delta J_s>=0$
for $1\leq s\leq n-1$.  Hence
$\omega_{{\cal M}}=-\sum_1^g<\delta J_{j+n}>\wedge\,\, dt_j$.
Further from \cite{ce,cf} one has 
$(\partial_j\psi^*/\psi^*)=-\lambda^j-\sum_1^{\infty}\sigma_s^j\lambda^
{-s}=-z^{-j}+\sum_1^{\infty}h^*_{s,j}z^s$ and a calculation gives
$\delta J_{j+n}(t)=\sum_1^{n-1}\delta J_sh^*_{n-s,j}$; hence
$\omega_{{\cal M}}=-Res_P(<\delta_t\psi^*\wedge \delta L_n\psi>/
<\psi^*\psi>)dp$ holds and an argument is suggested in \cite{kg}
to extend $\delta_t$ to a full $\delta$ so that the first formula
in (\ref{GK}) is proved.  For the second formula in (\ref{GK}) one looks
at $\delta log\psi^*=
\delta\left(c(t_i,\,i\geq 2)+\int_{x_0}^x\partial_xlog
\psi^*\right)=
\delta\sum_1^{\infty}\left(c_s(t_i,\,i\geq 2)+\int_{x_0}^xh_s^*z^sdx
\right)$.  This implies 
$-Res_P(<(\delta log\psi^*)\wedge (\psi^*\delta L_n\psi)>)/<\psi^*\psi>)=
\omega_{{\cal M}}=-\sum_1^{n-1}<\delta J_s\wedge\int_{x_0}^x\delta^*
h^*_{n-s}dx>$.
By definitions $\delta J_s$ does not contain variations of derivatives
of $u_i$ so we can write
$\delta J_s=-n(\partial h_s/\partial u)\delta u$ 
and this gives the second identity in (\ref{GK}).  We note here that there
is a little interplay between $\delta\sim\epsilon\partial_X$ and $X=
\epsilon x$ in the last two lines.  The examples in \cite{kg} have some
curious features but in any event one should keep in mind the conditions
$\delta H_s=0$ when selecting functions.  One notes that 
of course the symplectic
structure involving $\delta h_s,\,\,\delta^* h_s^*$, etc. does not
include the time dynamics so KdV $\sim L_2=\partial^2+u$ for example
and Bousinesq $\sim L_3=\partial^3+u\partial+v$.  Adaptions and 
applications of the above fibration framework to $SU(n),\,\,N=2$ susy
YM theory are also given in \cite{kg}.
\\[3mm]

\end{document}